\documentclass[twocolumn,showpacs,preprintnumbers,amsmath,amssymb]{revtex4-1}%
\usepackage{natbib}%
\usepackage{graphicx}
\usepackage{dcolumn}
\usepackage{bm}
\usepackage{color}%
\usepackage{amsmath}%
\usepackage{amsfonts}%
\usepackage{psfrag}%
\usepackage{fancyhdr}%
\usepackage{float}%

\usepackage{revsymb4-1}%

\begin{document}%



\title{%
Do magnetic fields enhance turbulence at low magnetic Reynolds number ?}
\author{%
Alban Poth\'erat$^{1,2}$ and Rico Klein$^1$}
\affiliation{%
$^1$Applied Mathematics Research Centre, Coventry University, Priory Street Coventry CV1 5FB, United Kingdom\\
$^2$ Laboratoire National des Champs Magn\'etiques Intenses- Grenoble, 
Universit\'e Grenoble-Alpes/CNRS, 25 Rue des Martyrs, B.P. 166 38042 Grenoble Cedex, France
}
\email{
alban.potherat@coventry.ac.uk}
\begin{abstract}%
Imposing a magnetic field on a turbulent flow of electrically conducting fluid incurs the Joule effect. A current paradigm is that the corresponding dissipation increases with the intensity of the magnetic field, and as a result
turbulent fluctuations are all the more damped as the magnetic field is strong.
While this idea finds apparent support in the phenomenology of
decaying turbulence, measurements of turbulence in duct flows
and other, more complex configurations have produced
seemingly contradicting results. The root of the controversy is that magnetic
fields promote sufficient
scale-dependent anisotropy to profoundly reorganise the structure of
turbulence, so their net effect cannot be
understood in terms of the additional dissipation only.
Here we show that when turbulence is forced
in a magnetic field that acts on turbulence itself rather than on the
mechanisms that generate it, the field promotes large,
nearly 2D structures capturing sufficient energy to
offset the loss due to Joule dissipation, with the net effect of
increasing the intensity of turbulent fluctuations.
This change of paradigm potentially carries important consequences for
systems as diverse as the liquid cores of planets, accretion disks and a
wide range of metallurgical and nuclear engineering applications.
\end{abstract}%

\keywords{%
magnetohydrodynamics, turbulence}
\maketitle
\begin{figure*}
\begin{center}
\begin{tabular}{ccc}
$l_z(l_\perp)\ll h$, $h_V=l_z(l_\perp)$&	
$l_z(l_\perp)\sim h$, $h_V=h$ &
$l_z(l_\perp)\gg h$, $h_V=h$\\
\hspace{-1cm}
\psfrag{Ic}{\textcolor{red}{$I_c$}}
\psfrag{Ic0}{\textcolor{red}{$I_c\simeq0$}}
\psfrag{Ib}{\textcolor{red}{$I_b$}}
\psfrag{It}{\textcolor{red}{$I_t$}}
\psfrag{Itb}{\textcolor{red}{$I_t=I_b$}}
\psfrag{I}{\textcolor{red}{$I_F$}}
\psfrag{I2}{\textcolor{red}{$I_F/2$}}
\psfrag{Uc}{\textcolor{blue}{$U$}}
\psfrag{Ub}{\textcolor{blue}{$U_b$}}
\psfrag{Ut}{\textcolor{blue}{$U_t$}}
\psfrag{Uh}{\textcolor{blue}{}}
\psfrag{Utb}{\textcolor{blue}{$U_t=U_b$}}
\psfrag{dh}{{$\delta_H$}}
\psfrag{h}{{$h$}}
\psfrag{lz}{{$l_z$}}
\psfrag{lp}{{$l_\perp$}}
\includegraphics[height=0.26\textwidth]{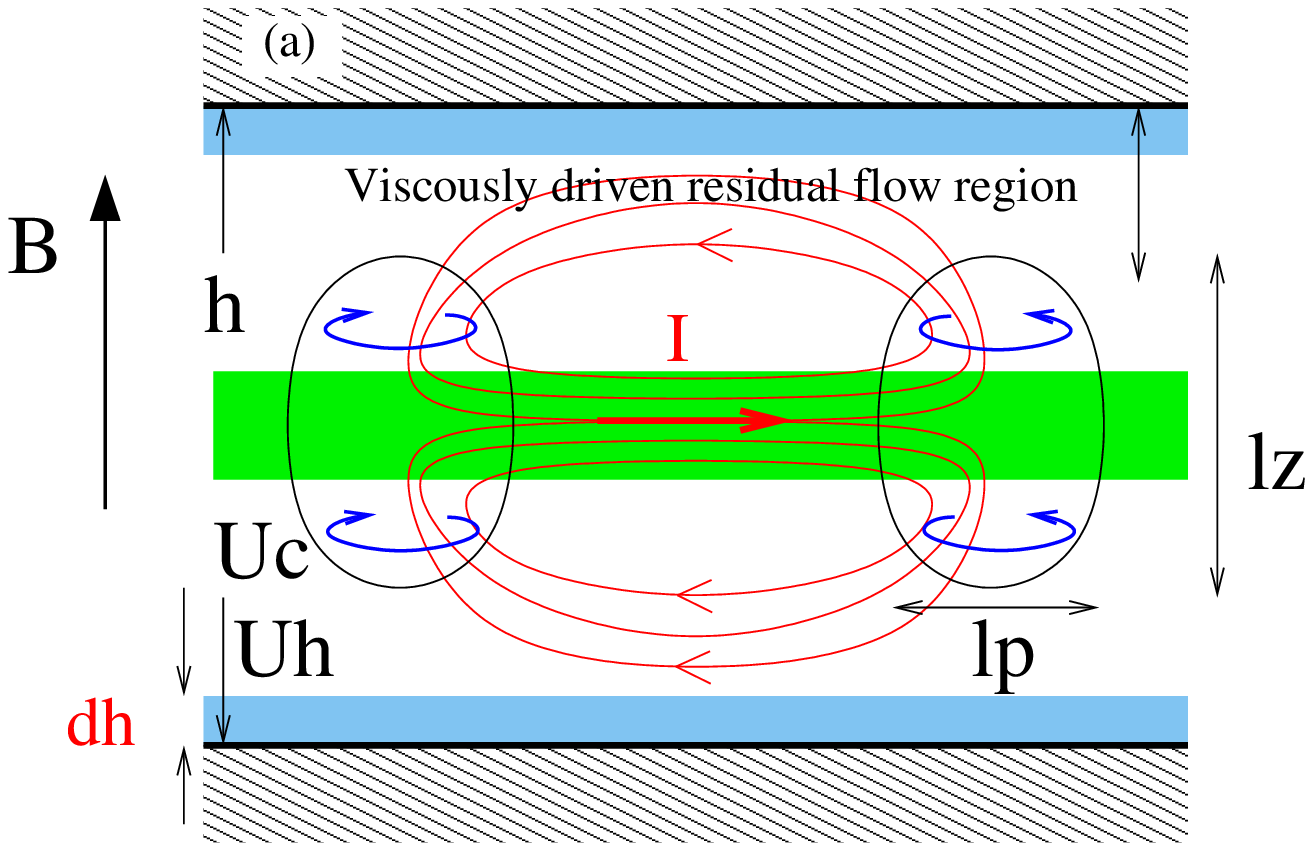}&
\hspace{-1.5cm}
\psfrag{Uh}{\textcolor{blue}{$U$}}
\psfrag{I}{\textcolor{red}{$I_F$}}
\includegraphics[height=0.26\textwidth]{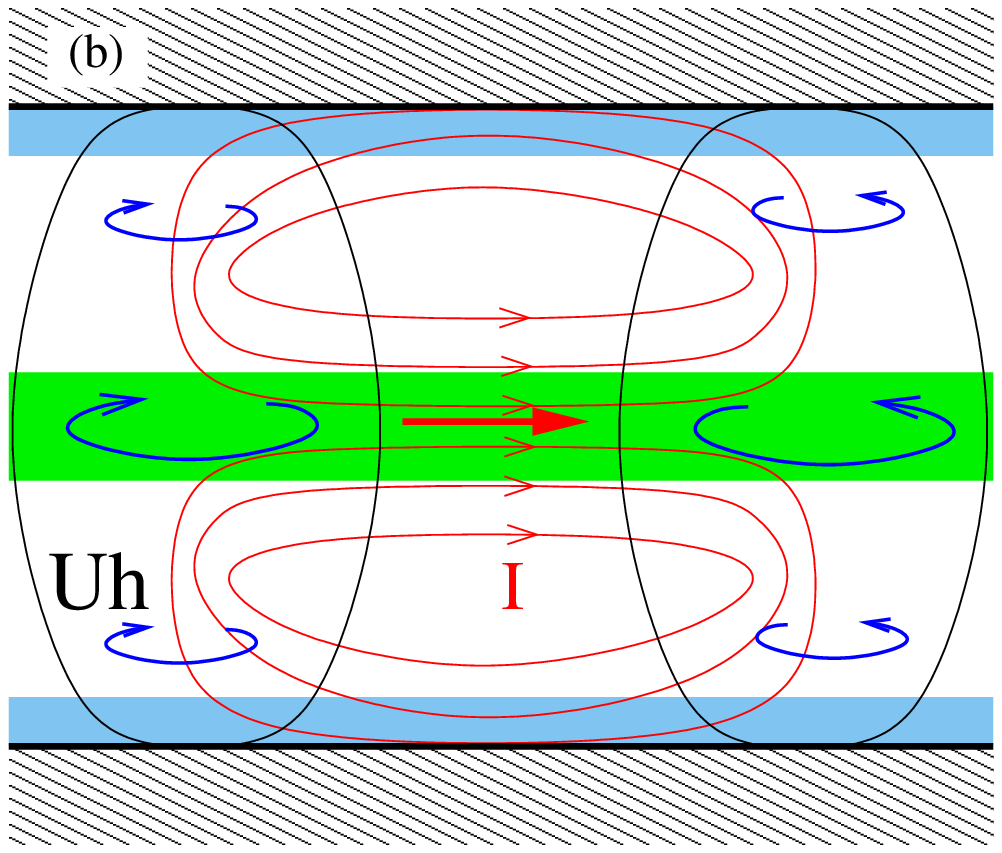}&
\hspace{-1.5cm}
\psfrag{Uh}{\textcolor{blue}{$U$}}
\psfrag{I2}{\textcolor{red}{$I_F/2$}}
\includegraphics[height=0.26\textwidth]{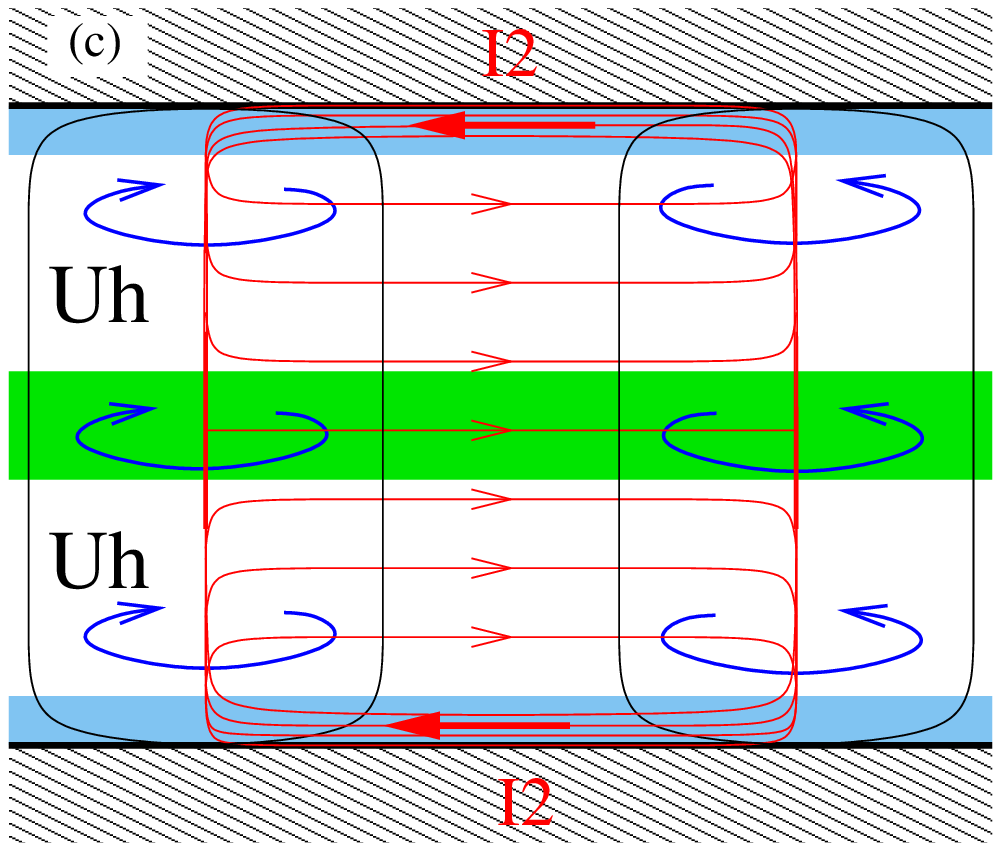}\\
\end{tabular}
\end{center}
\caption{Generic flow structures of lengthscale $l_\perp$ driven by 
an external force $\mathbf F$ (in the green area) in an {external} 
magnetic field $\mathbf B$.  
The Lorentz force diffuses momentum along the field over $l_z(l_\perp)$, and 
induces electric currents (in red). (a) {$l_z(l_\perp)\ll h$}: the current driven by the 
external forcing returns in the bulk. Only a residual flow exists near the 
walls. (b) {$l_z(l_\perp)\sim h$}: the current returns in the bulk and the Hartmann layers (in 
blue): the flow is still 3D but influenced by the walls. 
{(c) $l_z(l_\perp)\gg h$: the current returns equally in the top and bottom Hartmann layers, and not in the bulk: the flow is quasi-2D.} 
}
\label{fig:3dsketches}
\end{figure*}

%
{\section{Introduction \label{sec:intro}}}
Turbulent flows are often exposed to magnetic fields, 
either externally applied, 
or self-generated. 
In strong mean fields, the induced Lorentz force 
alters the way flows transport heat or mass, and dissipate energy \cite{shebalin83_jpp}.
Among the vast array of processes that are concerned are the dynamics of liquid planetary cores \cite{ryan2007_grl}, the solidification of metallic alloys \cite{prescott1995_jht,li2007_am},
and the cooling of nuclear reactors \cite{smolentsev2008_fed}.
The current paradigm is that the action of the field on the flow is a damping 
one, because of the extra dissipation incurred by the Lorentz force 
\cite{davidson1999_arfm}. It finds support in the phenomenology of freely 
decaying magnetohydrodynamic (MHD) turbulence, which decays faster under a higher 
externally imposed magnetic field \cite{moffatt67_jfm, kop15_jfm}. 
{Nevertheless, the presence of an external magnetic field does not, in 
general, damp instabilities or turbulent fluctuations. Quite the opposite 
happens in axisymmetric flows subject to axial and toroidal magnetic fields, 
where the magnetorotational and Tayler instabilities occur (MRI and TI, 
respectively). Both instabilities have been theoretically predicted from linear 
stability analysis \cite{tayler1973_mnras,liu2006_pre} but only recently observed in liquid 
metal experiments \cite{stefani2006_prl,seilmayer2012_prl}. MRI might explain 
why accretion disks, which should remain laminar in the absence of a magnetic 
field are in fact turbulent. 
TI is a good candidate to explain the mechanisms behind the Sun's dynamo \cite{balbus1998_rmp}. 
In both cases, the instability mechanisms involve a 
two-way coupling between the magnetic field and the flow.
Yet, enhancement of turbulent fluctuations has been reported in a number of 
examples where the magnetic field was imposed, and not influenced by the flow too: in numerical simulations and laboratory models for the continuous steel casting process, the complex 
DC magnetic field applied to control the flow was found to drive high 
amplitude turbulent fluctuations \cite{timmel2011_mmtb,chaudhary2012_mmtb,miao2012_mmtb}.
The picture is more complex in duct flows subject to an imposed magnetic 
field: when the flow is driven by pressure alone,} the intensity of 
turbulent fluctuations decreases in the central part of the duct at higher 
magnetic fields \cite{krasnov2012_jfm}. By contrast, earlier experiments 
clearly show an increase in turbulent intensity with an external magnetic field 
in duct flows where turbulence is generated by a grid 
\cite{kolesnikov74,eckert2001_ijhff,sukoriansky1986_ef}, {or in 
numerically forced flows in periodic domains \cite{boeck08_prl}}. 
In pressure-driven 
duct flows, turbulence is produced by shear layer instabilities, 
which are severely damped by the magnetic field, so \cite{krasnov2012_jfm} 
arguably confirms that the magnetic field damps this particular mechanism of 
turbulence production \cite{krasnov2010_jfm}.
However, the question of the effect of magnetic fields on turbulence itself 
remains open and needs to be addressed in conditions where it is not overshadowed by 
the magnetic field's influence on the mechanism forcing turbulence.\\
We tackle this problem by considering a generic plane channel geometry 
pervaded by a uniform magnetic field perpendicular to it. We assume that (i) that 
the flow is driven by an external force that is not affected by the 
magnetic field and (ii) { the magnetic field induced by the flow 
is sufficiently small for the externally applied magnetic field $\mathbf B$ to 
be considered constant, \emph{i.e.} that 
the magnetic Reynolds number is small 
\cite{roberts67}. This approximation applies to most laboratory experiments 
with flows of moderate intensity.}
The key mechanism at play in these conditions was first characterised by the authors of Ref. \cite{sm82}, who showed that in inducing eddy currents to oppose velocity 
gradients along the magnetic field, the Lorentz force diffuses momentum along $\mathbf B$ over a timescale $\tau=\tau_J(l_z/l_\perp)^2$. Here, 
$\tau_J=\rho/(\sigma B^2)$ is the Joule dissipation time, $l_\perp$ and $l_z$ are lengthscales across and along the field, $\sigma$ and $\rho$ are the fluid's 
electrical conductivity and density.
 In turbulent flows, diffusion is disrupted by inertial transfer acting over a 
structure turnover time $\tau_U\sim l_\perp/U(l_\perp)$, where 
$U(l_\perp)$ is a typical velocity at scale $l_\perp$. The competition 
between both processes determines the scale-dependent anisotropy 
as $l_z(l_\perp)\sim l_\perp N(l_\perp)^{1/2}$, $N(l_\perp)=\tau_U/\tau_J$ being the 
interaction parameter at scale $l_\perp$ (see the theory from Ref. \cite{sm82} and experimental evidence in Ref. \cite{pk14_jfm}). As a result of this 
process, velocity gradients are reduced under the action of the magnetic field 
as $\partial_z \sim l_z^{-1}\propto B^{-1}$, and so is the current density.
 In a channel of width $h$, perpendicular to the magnetic field, the flow dimensionality is therefore determined by the scale-dependent ratio
$l_z(l_\perp)/h$ \cite{pk14_jfm}: if $l_z(l_\perp)/h<1$ a structure of size $l_\perp$ is 3D. In the limit $l_z(l_\perp)/h\rightarrow\infty$, it is quasi-2D.
In the quasi-2D limit, no eddy currents remain in the bulk, because diffusion 
then takes place across the entire channel, except in Hartmann boundary layers along the channel walls \cite{moreau90}. {Current that would be directly induced in the bulk by an external force (through Lenz's law) returns through the Hartmann layers too. All three cases are illustrated in Fig. \ref{fig:3dsketches}.}\\
To understand how this effect determines the intensity of turbulence, we shall first translate this phenomenology into a scaling 
linking the relative turbulent intensity with the forcing and magnetic field 
intensities (section \ref{sec:theory}).
Second, we shall experimentally drive turbulence in a liquid metal experiment 
where the forcing intensity can be precisely set, and measure both relative 
and absolute turbulent intensities (section \ref{sec:exp}).\\

%
%
%
%
%
\section{Theory \label{sec:theory}}
\subsection{Scaling for the local, event-averaged Reynolds number.\label{sec:scalre}}
We consider the generic configuration of an electrically conducting fluid (density $\rho$, electric conductivity $\sigma$, and viscosity $\nu$) confined between 
two parallel walls distant by $h$ and pervaded by a uniform magnetic 
field $\mathbf B=B\mathbf e_z$ normal to them. 
The flow is driven by an external force density field 
$\mathbf F$ non-dimensionally measured 
by the Grashof number $\mathcal G=F l_\perp^3/(\rho\nu^2)$, where $l_\perp$ is a lengthscale perpendicular to $\mathbf e_z$ (Fig. \ref{fig:3dsketches}).
For simplicity, $\mathbf F$ is chosen normal to $\mathbf e_z$. 
The first step is to seek a scaling for the average flow intensity for a given 
external force and magnetic field.
Away from the wall, $\mathbf F$ is balanced by inertia and the Lorentz force.
At low magnetic Reynolds numbers, magnetic field fluctuations are 
negligible compared to the externally imposed field \cite{roberts67}; the  
governing equations can be expressed in the form of the incompressible 
Navier-Stokes equations with an added term representing the Lorentz force 
density $\mathbf J\times\mathbf B$ 
and the constraints that the velocity field $\mathbf u$ and current density 
field $\mathbf J$ must be solenoidal to ensure mass and charge conservation. 
\begin{eqnarray}
(\partial_t +\mathbf u\cdot\nabla) \mathbf u+\frac1\rho{\nabla p} &=&
\nu \Delta \mathbf u + \frac1\rho\mathbf J\times\mathbf B+\frac1\rho\mathbf F,\\
\nabla\cdot\mathbf u=0, && \nabla\cdot\mathbf J=0.
\end{eqnarray}
The system is closed with the addition of Ohm's law 
\begin{equation}
\frac1\sigma\mathbf J=-\nabla\phi +\mathbf u\times\mathbf B,
\end{equation}
where $\phi$ is the electric potential. Away from the walls, the curl of the 
Navier-Stokes equations and charge conservation yield the gradient of current 
density $J_z$ along $z$ \cite{pk14_jfm}:
\begin{equation}
-\partial_z J_z=\frac1{B}\left[-\rho\nabla \times \left(\mathbf u\cdot\nabla\right) \mathbf u + \nabla\times\mathbf F\right]\cdot \mathbf e_z.
\label{eq:curlns}
\end{equation}
Equation (\ref{eq:curlns}) expresses that horizontally divergent
eddy currents (\emph{lhs.}) are induced by the rotational part of the forcing 
(second term, \emph{rhs.}). Part of them returns through 
the bulk where the Lorentz force they induce balances inertia (first term, 
\emph{rhs.}). {Consider a  structure
of size $l_\perp$ and velocity $U$ extending over height $h_V$ along the
magnetic field. From Eq. (\ref{eq:curlns}) the total current generated by the forcing inside the structure and following this path scales as $I_B\sim(\pi \rho l_\perp^2 h_V/B)(U^2/l_\perp^2)$, and the current induced by the forcing scales as $I_F\sim(\pi \rho l_\perp^2 h_V/B)(F/\rho l_\perp)$.
The remaining current $I_H$ (\emph{lhs.} of (\ref{eq:curlns})) returns through 
the thin Hartmann boundary layers that develop along the walls \cite{moreau90}. 
The Lorentz force they generate there is opposed by
viscous friction only  (see Fig. \ref{fig:3dsketches}). This balance 
determines the thickness $\delta_H=B^{-1}(\rho\nu/\sigma)^{1/2}$ of these 
layers and the current density flowing through them $J_H\sim \sigma B U$ 
\cite{moreau90}. Consequently, the total current through the Hartmann layer 
scales as $I_H\sim 2\pi l_\perp J_H\delta_H$. From this phenomenology, 
Eq. (\ref{eq:curlns}) reduces to the conservation of the global current 
induced by the forcing $I_F=I_H+I_B$, and from the} scalings for 
each term, Eq.(\ref{eq:curlns}) is expressed non-dimensionally as
\begin{equation}
\mathcal G-Re^2\sim2\frac{h}{h_V}\left(\frac{l_\perp}h\right)^2 H\!a Re,
\label{eq:g-re}
\end{equation}
where $Re=Ul_\perp/nu$ is the Reynolds number,
and the Hartmann number $Ha=h/\delta_H$ provides a non-dimensional measure of 
$\mathbf B$.\\ 
The height $h_V$ of the structure in the bulk is determined 
by the anisotropic action of the Lorentz force as follows: 
In the limit $Re/H\!a\gg1$, the bulk of the flow is 3D, and 
$h_V$ is set by the momentum diffusion length defined in the introduction 
$h_V\sim l_z(l_\perp)$. 
In this case, inertia in the bulk consumes the larger part of the current 
induced by external forcing, so it remains from (\ref{eq:g-re}) that
\begin{equation}
Re\sim \mathcal G^{1/2}.
\label{eq:g-re_inert}
\end{equation}
Since the contribution of the Hartmann layers in (\ref{eq:g-re}) is small 
in this limit, Eq. (\ref{eq:g-re_inert}) remains valid 
whether or not the flow interacts with the walls (in this latter case the current induced by the forcing returns entirely over height $l_z$, determined by the balance between the 
Lorentz force and inertia (\emph{i.e.} $I_H=0$, as in Fig. \ref{fig:3dsketches}-(a)).\\
{
Experiments on electrically driven turbulence (See Fig. 4-a in Ref. \cite{pk14_jfm})
showed that this relation holds to a great precision for turbulent flows,
if the Reynolds number $Re=\langle|\mathbf u|\rangle L_f/\nu$ is expressed
in terms of the average velocity in the forced region and the forcing
lengthscale $L_f$ {(\emph{i.e} by setting $l_\perp=L_f$ and $U=\langle|\mathbf u|\rangle$ in the derivation above)}. Even for $Ha$ up
to $2\times10^4$, the left hand side correction in (\ref{eq:g-re}) due to
current circulating in the Hartmann layers remained small, which confirms the
validity of (\ref{eq:g-re_inert}).\\
}
\subsection{Scaling relation for the relative turbulent intensity.\label{sec:scaling_alpha}} 
We shall now seek an estimate for the turbulent intensity, for given control 
parameters $\mathcal G$ and $Ha$, \emph{i.e} for given forcing and magnetic 
fields. For this, we first note that unlike in hydrodynamic turbulence, 
dissipation in MHD turbulence is Ohmic {and roughly scales as $\epsilon_J\sim (\nu/h^2) Ha^2\|\mathbf u\|^2 (l_\perp/l_z)^2$ ( a more precise scaling will be obtained in this section). Consequently, it preferentially affects large 
scales exhibiting three-dimensionality (\emph{i.e} for which $l_z$ remains 
finite)}.
 Hence, 
 the tendency toward two-dimensionality of MHD turbulence, {which 
is a by-product of the dissipative process}, is independent 
of its spectral structure \cite{gallet2015_jfm} and the global dissipation 
can be captured by means of a power budget 
over a volume $V$ of height $h$ along $\mathbf B$ and size $l_\perp$ greater 
than the forcing scale. {Decomposing the velocity field $\mathbf u=\langle \mathbf u\rangle +\langle\mathbf u^{\prime 2}\rangle^{1/2}$ into average and fluctuations, we shall write this budget separating the current density field $\mathbf J$ into its average and fluctuating parts returning through the 
Hartmann layers $\langle\mathbf J_{2D}\rangle+\mathbf J'_{2D}$ and its average and fluctuating
parts returning through the bulk $\langle\mathbf J_{3D}\rangle+\mathbf J'_{3D}$:
\begin{eqnarray}
\langle
\int_V\mathbf F\cdot\langle\mathbf u\rangle+\left(\langle\mathbf J_{3D}\rangle+\mathbf J'_{3D}+\langle\mathbf J_{2D}\rangle+\mathbf J'_{2D}\right) &\times& \mathbf B \cdot \left(\langle\mathbf u\rangle + \mathbf u^\prime \right)\label{eq:power_sup} \nonumber \\
+\rho\nu \left(\langle\mathbf u\rangle+\mathbf u^\prime\right) \cdot \Delta \left(\langle\mathbf u\rangle+\mathbf u^\prime\right) dV\rangle &=& 0,
\end{eqnarray}
This way,} the budget reflects the structure of the eddy currents identified previously: 
The power $\mathcal P_F$ generated by the 
forcing is consumed in three ways: Joule dissipation due to current 
returning through the bulk $\epsilon_{3D}$, Joule dissipation due to currents 
returning through the Hartmann layers $\epsilon_{2D}$,  and viscous dissipation 
$\epsilon_\nu$ almost entirely produced in the Hartmann layers:
\begin{eqnarray}
\mathcal P_F+\epsilon_J^{3D}+\epsilon_J^{2D}+\epsilon_\nu=0.
\label{eq:power}
\end{eqnarray}
The contribution from inertial and pressure terms is 
neglected on the grounds that it is $N(l_\perp)$ times smaller than Joule
dissipation and that for strong enough magnetic fields, $N(l_\perp)\gg1$. This 
condition sets a minimum lengthscale $l_\perp$ for Eq. (\ref{eq:power}) to be 
valid over $V(l_\perp)$. 
{The forcing power is evaluated by noticing that eddy currents induced by the forcing diffuse over the lengthscale $l_z(l_\perp)$ built on the average velocity:
\begin{equation}
\mathcal P_F\sim l_\perp^2 l_z(l_\perp)|\langle \mathbf u \rangle||\mathbf F|\sim \rho \nu h \langle \mathbf u \rangle^2 {\mathcal G H\!a}{Re^{-3/2}}.
\label{eq:pi_sup}
\end{equation}
Joule dissipation in the bulk can be related to the velocity field by virtue of the expression of the rotational part of Lorentz force put forward by \cite{sm82}:
\begin{equation}
[\mathbf J\times\mathbf B]_{\nabla\times}=-\frac\rho{\tau_J} \Delta^{-1}\partial_{zz}^2 \mathbf u.
\end{equation}
Since bulk velocity gradients along the magnetic fields are controlled by 
the diffusion of momentum by the Lorentz force, $\Delta^{-1}\partial_{zz}^2\sim (l_\perp/l_z)^2$. Importantly, however,  
mean flow and fluctuations may significantly differ in intensity and may 
therefore diffuse over different lengths under the action of the Lorentz force, 
respectively  $l_z=l_\perp N(|\langle \mathbf u\rangle|)^{1/2}$ and 
$l_z^\prime=l_\perp N(\langle \mathbf u^{\prime 2}\rangle^{1/2})^{1/2}$. 
Hence, separating average and fluctuating parts yields:
\begin{eqnarray}
\epsilon_J^{3D}&=&
-\frac\rho{\tau_J}\int_V\left[ \langle\mathbf u\rangle\cdot\Delta^{-1}\partial_{zz}^2\langle\mathbf u\rangle+\langle\mathbf u^\prime\cdot\Delta^{-1}\partial_{zz}^2\mathbf u^\prime\rangle\right]dV,\nonumber\\
&\sim& \rho \nu h \langle \mathbf u \rangle^2 Re(1+\alpha^3), \label{eq:eps3d_sup}
\end{eqnarray}
where 
\begin{equation}
\alpha=\langle\mathbf u^{\prime 2}\rangle^{1/2}/|\langle \mathbf u\rangle|
\end{equation}
is the relative turbulent intensity.\\
Dissipation in the Hartmann layers is equally Ohmic and viscous
$\epsilon_J^{2D} =\int_V (\langle\mathbf J_{2D}\rangle+\mathbf J_{2D}^\prime) \times \mathbf B \cdot \mathbf u dV\simeq\epsilon_\nu=\int_V \rho\nu\mathbf u\cdot\Delta \mathbf u dV\simeq\int_V \rho\nu\mathbf u\cdot\partial^2_{zz} \mathbf u dV$ and is estimated from the thickness of the
Hartmann layer $\delta_H=h/Ha$ \cite{moreau90}:
%
\begin{equation}
\epsilon_J^{2D}\simeq\epsilon_\nu \sim-
\rho \nu h \langle\mathbf u \rangle^2 \left(\frac{l_\perp}{h}\right)^2H\!a(1+\alpha^2).
\label{eq:eps2d_sup}
\end{equation}
By virtue of (\ref{eq:pi_sup},\ref{eq:eps3d_sup},\ref{eq:eps2d_sup}), and
(\ref{eq:power_sup}) becomes
\begin{equation}
\mathcal G {H\!a}{Re^{-1/2}}\left(\frac{l_\perp}{h}\right) \sim Re^2(1+\alpha^3)+2 H\!a Re\left(\frac{l_\perp}{h}\right)^2(1+\alpha^2).
\end{equation}
Lastly, $Re$, which is not known \emph{a priori}, is expressed  in terms of
control parameter $\mathcal G$ through scaling (\ref{eq:g-re_inert}). This 
yields a simplified relation linking the relative turbulent intensity $\alpha$ to control parameters $H\!a$ and $\mathcal G$:
}
\begin{equation}
H\!a \mathcal G^{-1/4}\left(\frac{l_\perp}{h}\right) \sim (1+\alpha^3)+ 2{H\!a}{\mathcal G^{-1/2}} \left(\frac{l_\perp}{h}\right)^2(1+\alpha^2).
\label{eq:alpha}
\end{equation}
These estimates reflect that increasing $B$ (or $Ha$) extends the thickness of the forced region by 
$l_z\propto B$.
Consequently, the 
forcing power increases linearly with $B$. On the other hand, the power 
dissipated in the bulk does not vary with $B$ because the intensity of the eddy currents there is governed by velocity gradients along $z$ that decrease as 
$l_z^{-1}\propto B^{-1}$, 
thus exactly compensating the increase in $B$. 
%
%
%
%
%
Plots of (\ref{eq:alpha}) in Fig. \ref{fig:turbu_int_rel}  show that the relative intensity of 
turbulence increases with the externally imposed magnetic field (or $H\!a$), 
with $\alpha\propto H\!a^{1/3}$ for $\alpha>1$.
The underlying phenomenology reflects anisotropic mechanisms that are the 
hallmark of MHD turbulence: in 3D flows, high velocity gradients along the 
magnetic field incur Joule dissipation that damps velocity fluctuations.
As the magnetic field increases, turbulent structures elongate as 
$l_z^\prime(l_\perp)$ increases, velocity gradients weaken, so does the Joule 
dissipation they incur, 
 and turbulent fluctuations retain more energy.
Through this process, the  relative intensity of turbulent fluctuations
 is higher for higher magnetic fields.\\ 

%
\section{Experiments\label{sec:exp}}
\subsection{Experimental approach.\label{sec:expmeth} }
Our second step is to test scaling relation (\ref{eq:alpha}) on the FLOWCUBE 
facility, which reproduces the generic configuration from 
Fig. \ref{fig:3dsketches} in a 10 cm-cubic vessel filled with liquid metal 
pervaded by the near-homogeneous magnetic field generated in the bore of 
a superconducting solenoidal magnet. {The metal is Gallinstan, 
an eutectic alloy of gallium, indium and tin that is liquid a room temperature.} Full details and validation of the experimental setup are provided in 
\cite{pk14_jfm,bpdd2017_ef}. 
An electrically generated force of lengthscale $L_f$ keeps turbulence in a 
statistically steady state as follows: 
 Electric current $I$ locally injected through one of the channel walls 
(arbitrarily the bottom wall) forces horizontally divergent currents 
through the bulk and the Hartmann layers. {Their interaction 
with the externally imposed magnetic field exerts} a driving Lorentz force
on the flow. This mechanism ensures that the forcing is directly controlled 
by the injected current and does not vary with the external magnetic field.
{This somewhat counter-intuitive result} comes as a consequence of the diffusion by the Lorentz force over 
$l_z\sim L_f N^{1/2}=L_f Ha/Re^{1/2} (L_f/h)$ of the current $I$ injected 
through the wall. {Because of this effect, the volume of forced fluid increases linearly with the magnetic field}. Hence, the horizontal current density driving fluid motion 
scales as $J_\perp\sim I/(2\pi L_f l_z)$. It follows that the force density 
driving turbulence is  controlled by the injected current $I$ only, as {$F\sim IB/(2\pi L_f l_z)=I(\rho\nu/\sigma)^{1/2}/(2\pi L_f^3)Re^{1/2}$,  and does not depend on the magnetic field}. By virtue of (\ref{eq:g-re_inert}), 
$\mathcal G= [I/(2\pi(\nu^3\sigma\rho)^{1/2})]^{{4/3}}$ is varied solely by 
adjusting the injected current $I$.
This method can be understood as an extension to 3D turbulence of the method of  
\cite{sommeria88} to force quasi-2D MHD flows, where the forcing was 
also controlled by the current measured non-dimensionally by the parameter 
\begin{equation}
Re_0=\frac{I}{2\pi (\sigma\rho\nu^3)^{1/2}}=\mathcal G^{3/4}.
\end{equation} 
Based on this principle, turbulence is driven in FLOWCUBE by injecting current 
through electrodes embedded flush in the bottom wall, arranged in a $10\times10$
 square array of step $L_f=1$ cm and alternately connected to the positive and 
negative poles of a DC current supply. The corresponding average flow is a 
crystal of columnar vortices attached to the bottom wall extending 
by $l_z(L_f)$ into the bulk through diffusion by the Lorentz force {(see 
\cite{kp10_prl, pk14_jfm} for a detailed analysis of this flow, and 
\cite{p12_epl,prcd13_epje,bpd2015_jfm} for a theory explaining how its 
componentality is controlled by the magnetic field)}.
The region just outside the bottom Hartmann layer is always in the forcing 
region and is representative of the bulk of forced turbulence. On the other 
hand, the region just outside of the Hartmann layer near the top wall may lay 
within the forcing region if $l_z(L_f)/h=N(L_f)^{1/2}L_f/h>1$ or outside it if 
$l_z(L_f)/h<1$. Since $L_f$ is the scale of the average flow, $Re$, $N$ and 
$\mathcal G$ are evaluated taking $l_\perp=L_f$. {As in Section \ref{sec:theory}, 
we shall distinguish between $N$ and $N'$, built 
on the average velocity and the fluctuations, respectively}.\\
Bulk velocities are measured locally just outside the bottom and top Hartmann 
layers by electric potential velocimetry  \cite{kljukin98_ef} 
through 192 electric potential probes fitted flush in each of  the Hartmann walls \cite{pk14_jfm}.
Average and RMS fluctuations near bottom and top walls
(denoted by indices $_b$ and $_t$ respectively) are obtained 
from spatial and time averages of time-dependent signals of 
$\mathbf u_{b,t}(x,y,t)$, {recorded in a statistically state}.\\
\begin{figure}
\includegraphics[width=0.5\textwidth]{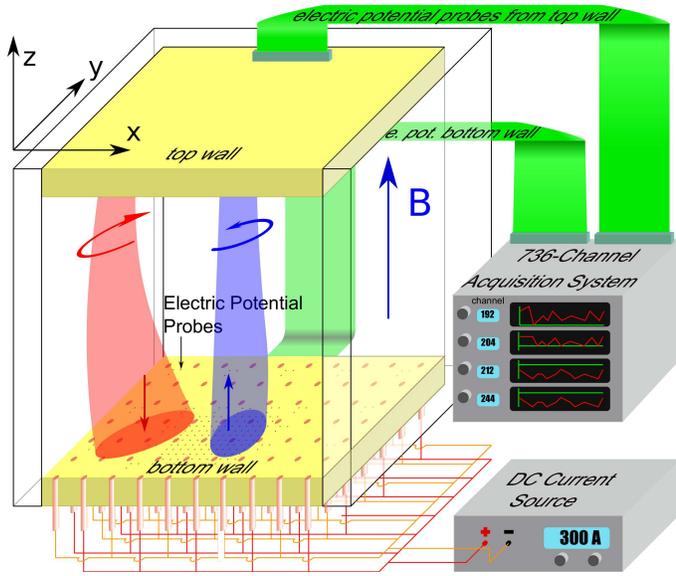}
\caption{Simplified sketch of the FLOWCUBE platform for the study of electrically driven liquid metal flows representing the main vessel filled with liquid metal with Hartmann walls (in yellow) and walls parallel to the magnetic field (transparent). The flow is forced by injecting DC electric current through a square array of electrodes embedded flush in the bottom wall. This creates a crystal of
columnar vortices, which in destabilising, gives rise to turbulence. An external magnetic field is applied by means of a large solenoidal cryomagnet (not represented). Depending on the intensities of the magnetic field and turbulence, turbulence extends to part of the bulk (if {$l_z'(L_f)<h$}) or up to the top wall (if {$l_z'(L_f)\gtrsim h$}). Velocities are reconstructed just outside the boundary layers near the top and bottom walls, from measurements of electric potential at 384 contact probes embedded flush in the top and bottom walls (black dots) and connected to a 768-channel, high precision acquisition system through printed circuit built-in the Hartmann walls. \label{fig:flowcube}}
\end{figure}
%
\begin{figure}
\begin{tabular}{r}
\includegraphics[width=0.49\textwidth,height=0.05\textwidth]{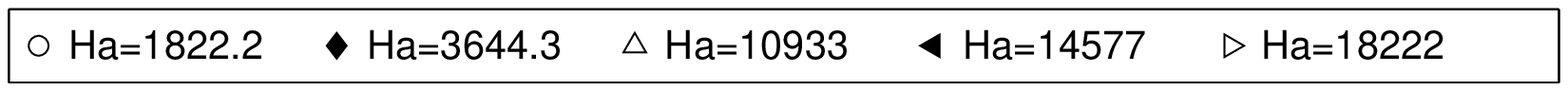}\\
\psfrag{GH}{$Re_0=\mathcal G^{3/4}$}
\includegraphics[width=0.49\textwidth]{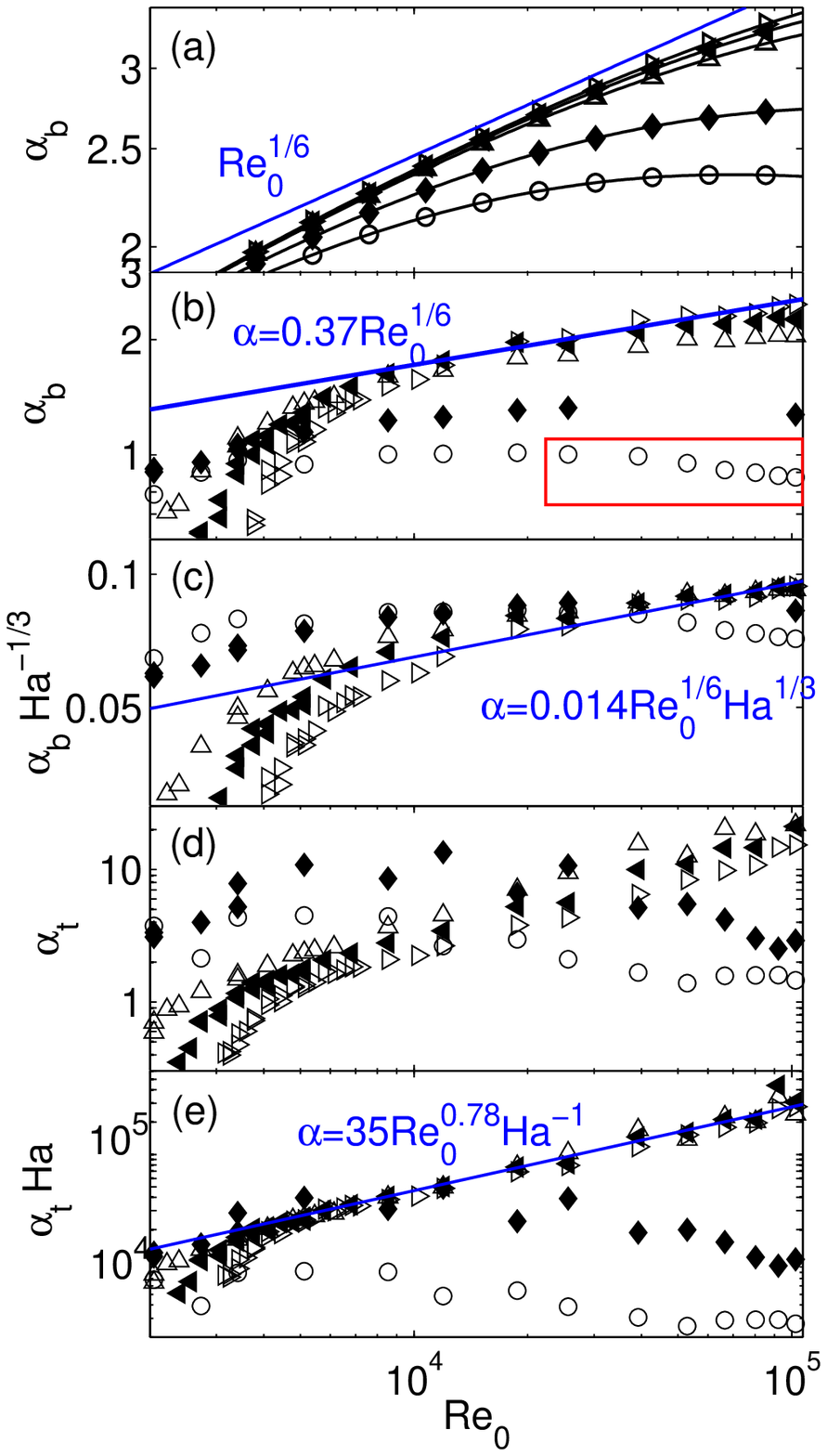}\\
\end{tabular}
\caption{Relative turbulent intensity near the bottom wall (forced region) (from (a) Eq. (\ref{eq:alpha}), (b,c): experiment) and (d) top wall, $\alpha_{b,t}=\langle \mathbf u_{b,t}^{\prime 2}\rangle^{1/2}/|\langle\mathbf u_{b,t}\rangle|$. The red rectangle indicates the regime of strongest three-dimensionality, where turbulent fluctuations increase with the external magnetic field and decrease as $\mathcal G^{-1/12}$ with the forcing. The slope $Re_0^{1/6}(\propto \mathcal G^{1/8})$ is indicative only, as the solution of (\ref{eq:alpha}) for $\alpha(\mathcal G)$ is not a power law.
\label{fig:turbu_int_rel}}
\end{figure}
\begin{figure}
\begin{tabular}{r}
\includegraphics[width=0.49\textwidth,height=0.05\textwidth]{legend.eps}\\
\vspace{-0.5cm}
\psfrag{GH}{$Re_0=\mathcal G^{3/4}$}
\includegraphics[width=0.49\textwidth]{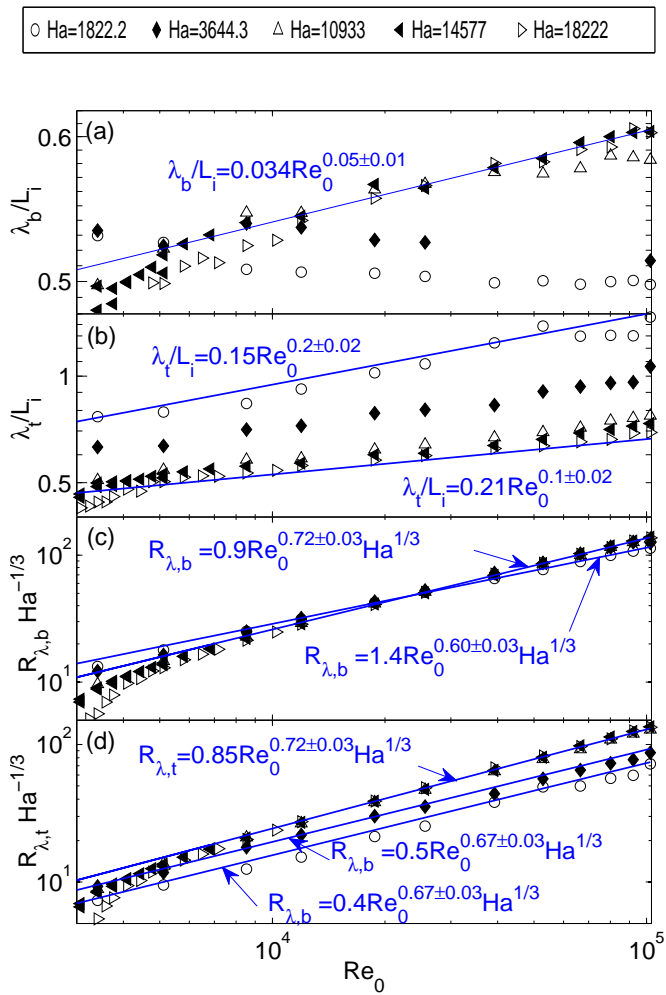}\\
\end{tabular}
\caption{
Taylor microscale near (a) the bottom wall ($\lambda_b$, forced region), and (b) the top wall ($\lambda_t $, outside the forced region if $l_z(L_f)>h$), and Reynolds number $R_\lambda$ based on (c) $\lambda_b$ and (d) $\lambda_t$, showing increasing absolute turbulent intensity with the external magnetic field both within and outside the forced region.
\label{fig:turbu_int_abs}}
\end{figure}
%

\subsection{Relative turbulent intensity. \label{sec:turbu_rel}} 
%
Relative turbulent intensities near the bottom and top walls 
$\alpha_{b,t}=\langle \mathbf u_{b,t}^{\prime 2}\rangle^{1/2}/|\langle\mathbf u_{b,t}\rangle|$ are 
shown in Fig. \ref{fig:turbu_int_rel}. 
Within the forced region of the flow (\emph{i.e.} near the bottom wall, see Fig.
 \ref{fig:turbu_int_rel}-b), $\alpha_b$ is found to always increase with $Ha$ 
and its variations are very well reproduced by the scaling relation (\ref{eq:alpha}): in both theory and experiments, $\alpha_b\propto Ha^{1/3}$ 
for $\alpha_b>1$, a regime where both the top and the bottom walls lie within 
the forced region (\emph{i.e} $l_z(L_f)>h$). When $\alpha_b\lesssim1$, the 
dependence on $Ha$ is even 
stronger than $Ha^{1/3}$. In this regime, $l_z(L_f)<h$ so the forcing region 
progressively extends across the channel as $Ha$ is increased. The increase in 
actual vortex length along $\mathbf B$ causes {a drastic 
reduction of velocity gradients along $z$ and a} drop in bulk dissipation 
that translates into the sharper increase of turbulent fluctuations observed.\\  
Equation (\ref{eq:alpha}) and its underlying phenomenology are further validated in 
the most 3D regimes ($Ha=1822$, high $\mathcal G$), where $\alpha_b$ decreases 
with $\mathcal G$ (Region inside the red rectangle in Fig. \ref{fig:turbu_int_rel}-b). This behaviour is captured in the limit where Joule dissipation mostly occurs in 
the bulk, 
in which case (\ref{eq:alpha}) reduces to $\alpha_b\propto \mathcal G^{-1/12} \propto Re_0^{-1/9}$.\\
%
By contrast Eq. (\ref{eq:alpha}) does not apply to relative fluctuations $\alpha_t$ 
near the top wall, which lays outside the forced region 
if $l_z(L_f)\lesssim h$. Indeed relative turbulent fluctuations there 
first sharply increase with $H\!a$
for $H\!a<7500$ as large scale fluctuations diffuse up when $H\!a$ increases, 
to progressively reach the top wall. 
Fluctuations then decrease slightly with $H\!a$: rather than a loss of 
intensity in turbulence, we shall see that this reflects an increasing 
intensity of the mean flow near the top wall.\\

\subsection{Absolute turbulent intensity.\label{sec:turb_abs}}
To isolate the influence of the magnetic field on turbulent fluctuations 
from that on the mean flow, 
absolute turbulent intensity was measured through the microscale Reynolds number 
\begin{equation}
R_{\lambda,b,t}=\langle \mathbf u_{b,t}^{\prime 2}\rangle^{1/2}\lambda_{b,t}/\nu,
\end{equation}
 based on the Taylor microscale 
\begin{equation}
\lambda_{b,t}=\left[\frac{\langle \mathbf u_{b,t}^\prime(\mathbf x)\cdot \mathbf u_{b,t}^\prime(\mathbf x+\mathbf r)\rangle}{\partial^2_{rr}\langle \mathbf u_{b,t}^\prime(\mathbf x)\cdot \mathbf u_{b,t}^\prime(\mathbf x+\mathbf r)\rangle}\right]^{1/2} (\mathbf r=0),
\end{equation}
 which is representative of the inertial range 
of turbulence \cite{frisch95}. 
{The variations of $\lambda$ near the bottom ($\lambda_b$, Fig.  \ref{fig:turbu_int_abs}-a) and top walls ($\lambda_t$, Fig.  \ref{fig:turbu_int_abs}-b)
 turn out to be very weak. 
Hence, the variations of $R_\lambda$ 
are driven by the velocity fluctuations. 
Just like the relative turbulent intensity, the microscale Reynolds number 
(Figs. \ref{fig:turbu_int_abs}-c and 
\ref{fig:turbu_int_abs}-d) follows a scaling of $R_{\lambda,b,t}\propto Ha^{1/3}$ in the 
limit of high $H\!a$. There are, however, two important differences. First,  
unlike relative fluctuations, absolute fluctuations 
 always increase with $H\!a$, both inside and outside the forcing 
region.  
Second, they always increase with the forcing too, confirming that the decrease in relative turbulent 
intensity with either the forcing (for low $H\!a$) or the magnetic field outside the forcing region 
reflects an intensification of the average flow, and not weakening turbulent fluctuations.}\\
%

\section{Discussion.\label{sec:discussion}} 
These results provide robust evidence that the 
intensity of forced MHD turbulence 
increases with an externally applied magnetic field,
{ when the forcing density is kept constant and when
the damping of energy transfer from mean flow to turbulence by the 
magnetic field does not mask its influence on turbulence itself.
In the FLOWCUBE setup, constant forcing density is obtained by driving 
turbulence with a constant electric current but the same result can be achieved 
with other types of forcing mechanisms: If the flow had been locally forced by 
means of a propeller, since the forced 
volume increases as $l_z(L_f)\propto B$, the total torque on the propeller 
 would have to increase  with $B$ to keep the force density constant 
over the entire forced volume. Regardless of the forcing mechanism, in any given 
volume of forced turbulence, velocity gradients along the field are smoothed out by the Lorentz force as the field increases. {This mechanism 
reduces Joule dissipation. It allows turbulent fluctuations to retain more 
energy, and thereby leads turbulent intensity to increase as the magnetic field 
is increased}. In grid turbulence experiments 
\cite{kolesnikov74,alemany79, eckert2001_ijhff,sukoriansky1986_ef}, the average fluid velocity relative to the grid was kept constant. 
This impled increasing the driving force with the magnetic field to overcome 
the obstacle resistance. Equation (\ref{eq:g-re_inert}) implies that 
keeping the average velocity constant is indeed 
equivalent to keeping the force density constant, so the phenomenology 
identified in the FLOWCUBE applies to these experiments on duct flows too. Indeed, \cite{eckert2001_ijhff} 
shows an increase of approximately $\alpha\propto B^{0.5\pm0.1}$ for 
$Ha\gtrsim 10^3$, which is consistent with our own findings at low values of $Ha$.
Hence, while magnetic fields may damp instabilities in duct flows, they enhance 
turbulent fluctuations in general, in duct flows and other configurations.  
This phenomenon stems from the promotion of larger, more two-dimensional scales at higher magnetic fields.
The higher the fields, the wider the range of quasi-2D scales and the more intense
the turbulent fluctuations. This mechanism is further reinforced 
as large quasi-2D scales promote an inverse energy cascade that channels 
the energy injected by the forcing to them 
\cite{reddy2014_pf,deusebio2014_pre,sommeria86}.\\
In our experiment, 
the forcing, rather than turbulent production itself was kept constant and 
independent of the field. Hence, the increase in turbulent intensity 
took place despite a possible loss in the energy transferred from the mean flow 
to turbulence. A more precise characterisation of turbulence enhancement 
would be obtained by holding turbulent production constant, which can 
be envisaged in numerical simulations, rather than experiments.}\\
An important point to notice is that the reduction of velocity gradients with 
increasing magnetic field is the core effect driving turbulence enhancement. 
As a direct consequence, this mechanism cannot act when both the average flow 
and fluctuations cease to exhibit any gradients along the magnetic field, 
\emph{i.e.} when all flow structures are quasi-2D {(as on Fig. \ref{fig:3dsketches}-(c))}. Such a regime is not reached 
in FLOWCUBE for the magnetic fields considered here. However, 
\cite{eckert2001_ijhff} precisely reports a decrease in turbulent intensity 
with the fluctuations in the quasi-2D regime, which provides further support 
of the turbulence enhancement scenario we put forward.\\
Finally, two-dimensionalisation and energy cascades 
toward large scales exist in other systems. {For example, 
rotation promotes less dissipative quasi-2D 
turbulence too,
as the Coriolis force plays a similar role to that of the Lorentz force in 
MHD turbulence \cite{cambon1997_jfm}}. Hence, it can be expected that the 
intensity of turbulent fluctuations should increase with rotation too, as 
unlike the Lorentz force, the Coriolis force generates no dissipation to 
oppose this mechanism.
{\acknowledgements
A. Poth\'erat acknowledges support from the Royal Society under the Wolfson 
Research Merit Award scheme (Grant No. WM140032) and the International 
Exchanges scheme (Grant No. IE140127), from Universit\'e 
Grenoble Alpes and from Grenoble INP for the invited Professor positions they have 
granted him during the course of this project.}

\bibliographystyle{plain}%
\setlength\bibsep{0pt}%
\bibliography{pk15.bib}%

\end{document}